\documentclass[twocolumn]{webofc}
\usepackage[varg]{txfonts}   
\begin{document}
\title{Survival Mediated Heavy Element Capture Cross Sections}
%
%
\author{\firstname{Walter Loveland} \lastname{}\inst{1}\ \and
        \firstname{Larry Yao} \lastname{}\inst{1}\fnsep\
}

\institute{Dept. of Chemistry, Oregon State University, Corvallis, OR 97331 USA}
\abstract{%
  Formally, the cross section for producing a heavy evaporation residue, 
$\sigma_{\rm EVR}$, in a fusion reaction can be written as
\begin{equation}
\sigma_{\rm EVR}(E)=\frac{\pi h^2}{2\mu E}\sum\limits_{\ell=0}^\infty
(2\ell+1)T(E,\ell)P_{\rm CN}(E,\ell)W_{\rm sur}(E,\ell),
\end{equation}
where $E$ is the center of mass energy, and $T$ is the probability of the 
colliding nuclei to overcome the potential barrier in the entrance channel 
and reach the contact point.  $P_{\rm CN}$ is the probability that the 
projectile-target system will evolve from the contact point to the 
compound nucleus. $W_{\rm sur}$ is the probability that the compound 
nucleus will decay to produce an evaporation residue rather than fissioning.  
However, one must remember that the 
$W_{\rm sur}$ term effectively sets the allowed values of the spin, which in turn, restricts the values of the capture and fusion cross sections. 
We point out the implications of this fact for capture cross sections for heavy element formation reactions.
}
\maketitle
\section{Introduction}
\label{intro}
  Formally, the cross section for producing a heavy evaporation residue, 
$\sigma_{\rm EVR}$, in a fusion reaction can be written as as the product of three factors, (see equation 1) that express the probability of bringing the colliding nuclei into contact, having that configuration evolve inside the fission saddle point and having the resulting nucleus survive against fission.

\label{intro2}

\begin{figure}[h]
\centering
\includegraphics[width=8cm]{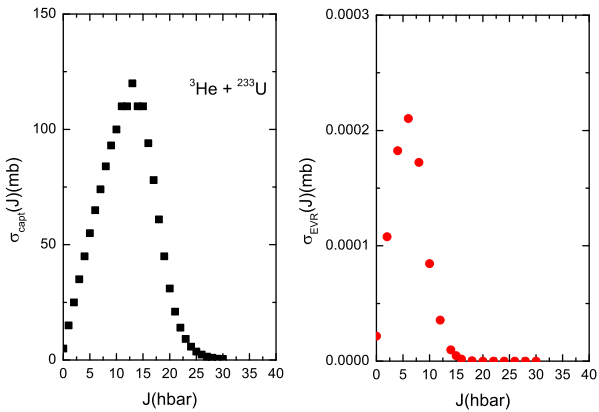}
\caption{Spin dependence of capture and evaporation residue formation cross sections for $^{3}$He + $^{233}$U where the lab energy of the $^{3}$He was 44.7 MeV.}
\label{fig-1}       
\end{figure}

Conventionally the EVR cross section is separated into three individual
reaction stages (capture, fusion, survival) motivated, in part, by the 
different time scales of the processes. However, one must remember that the 
$W_{\rm sur}$ term effectively sets the allowed values of the spin, which in turn, restricts the values of the capture and fusion cross sections. 
To illustrate this point, we show, in Figure 1, the spin dependence of capture cross section for the reaction of $^{3}$He + $^{233}$U and the spin dependence 
of the evaporation residue cross section for the 5n channel \cite{Carolla}.  Note the restrictions on the surviving spins due to the fission probability and the resulting differences between the capture and evaporation residue cross sections.

\section{Survival Probabilities}

For the most part, the formalism for calculating the survival, against fission, of a highly excited nucleus is understood \cite{Zaggy}.  One begins with a single particle model \cite{Ig83} of the level density in which one allows the level density parameter to be a function of the excitation energy.  Masses and shell corrections are taken from \cite{Mol16}.  The deformation dependent collective enhancement of the level density is taken from \cite{Zaggy01} for a particular xn outgoing reaction channel.  Standard formulas are used to calculate the decay widths for decay by neutron, charged particle and $\gamma$-emission.  The fission width is corrected for Kramers effects \cite{Kr40}.  The fission barrier heights are calculated using liquid drop barriers and excitation energy dependent shell corrections.

What are the uncertainties in these calculations of W$_{sur}$?   Lu and Boilley \cite{lb} found collective enhancement effects, Kramers effects, and the overall fission barrier height to have the biggest effect on the calculated survival probabilities.  Loveland \cite{wdl} concluded that, in general, fission barrier heights are known to within 0.5 - 1.0 MeV.  For super-heavy nuclei, the change of fission barrier height by 1 MeV in each neutron evaporation step can cause an order of magnitude uncertainty in the 4n-channel. To minimize sensitivity to this factor, we have restricted our attention to nuclei where Z $\leq$ 110 where the barrier heights are better known.

\section{Outline of Calculations}

We begin with the compilation of Duellmann of evaluated evaporation residue cross sections for reactions that produce heavy nuclei \cite{Chris}.  This compilation involves $\sim$ 500 different reactions producing nuclei from Z=80 to Z=122, although we have limited our calculations to cases where Z$_{CN}$ $\leq$ 110.  For each reaction (projectile, target and beam energy) we calculate the spin dependent capture cross sections using the Bass model \cite{oleg}. The beam energies are chosen to correspond to the maximum of the excitation functions for the reaction channel being studied.  For each value of the angular momentum, $\ell$, the survival probability is calculated using the formalism(s) of \cite{Zaggy}.  If we assume P$_{CN}$ is one, then, according to equation 1, we can  straightforwardly estimate the evaporation residue formation cross section.  If the calculated evaporation residue cross section is significantly greater then the measured cross section for this reaction, we have evidence, perhaps quantitative,  that P$_{CN}$ is less than 1.  

As a ``reality check" on these results, we calculated the EVR spin distribution for the reaction $^{176}$Yb($^{48}$Ca,4n)$^{220}$Th and compared it to the measured spin distribution of Henning et al. \cite{greg} (Figure 2) for the same reaction. The comparison of the results for the measured and calculated spin distributions indicates that the calculational procedure used in this work is appropriate. This same sort of agreement between the calculated spin distributions and the measured distributions was observed by \cite{Zaggy01} for the $^{208}$Pb($^{48}$Ca, 2n)$^{254}$No reaction. 

\begin{figure}[h]
\centering
\includegraphics[width=8cm]{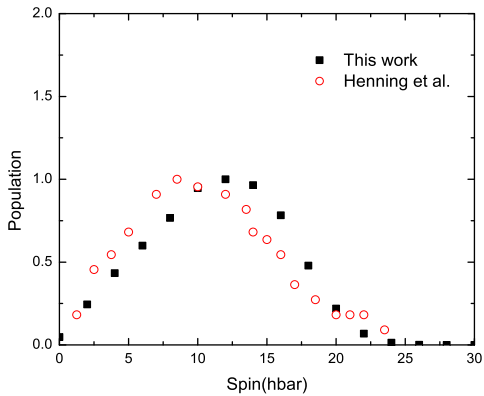}
\caption{Spin dependence of calculated and measured evaporation residue formation cross sections for the  $^{176}$Yb($^{48}$Ca,4n)$^{220}$Th reaction.}
\label{fig-2}       
\end{figure}

\subsection{Limitations of Calculational Procedure}
We have used relatively simple tools (the semi-empirical Bass model  \cite{oleg}) to simulate the dynamics of the capture process.  Better tools might have included coupled channel calculations for the capture process, taking into account the deformations of the target and projectile nuclei and the excitations of collective modes. The procedures used to calculate the survival probabilities are thought to reflect all the important features of these processes.  The success in calculating the EVR spin distributions (Fig. 2 and \cite{Zaggy01} support the methods chosen.  The dramatic decrease in the survival probabilities for higher J values (as shown in Figures 3-4) demonstrate the important qualitative effect being studied in this work.
\section{Results of Calculations}

\subsection {Z$_{1}$Z$_{2}$  <  700}

There are 76 cases in this category.  In each case, P$_{CN}$ was calculated to be 1, i.e., the calculated evaporation residue cross section agreed with the measured cross section within experimental error.  A sampling of some typical cases for Z$_{CN}$ = 94-98, 101-103 and 104-105 is shown in Figure 3.

\begin{figure}[h]
\centering
\includegraphics[width=8cm]{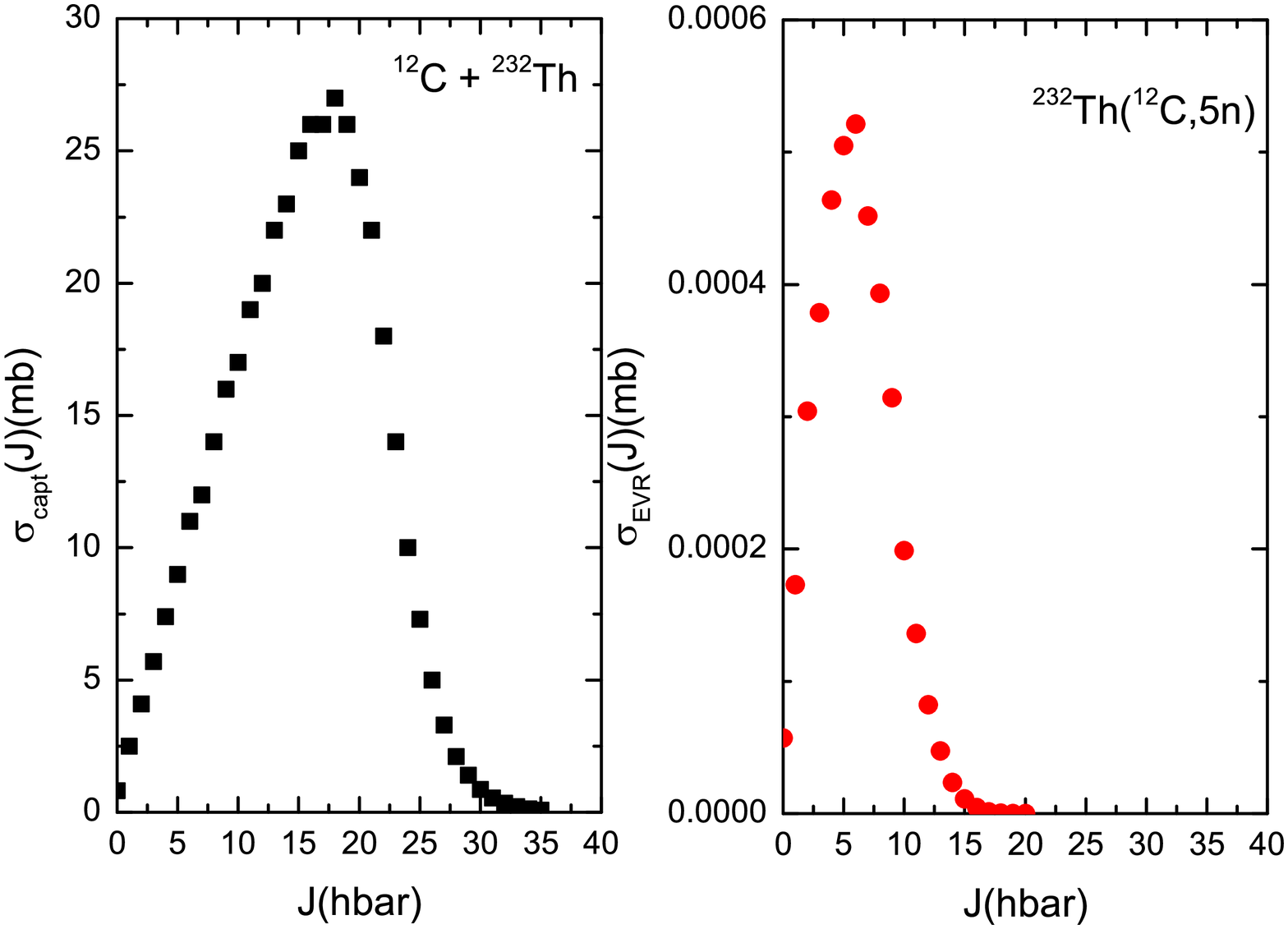}
\includegraphics[width=8cm]{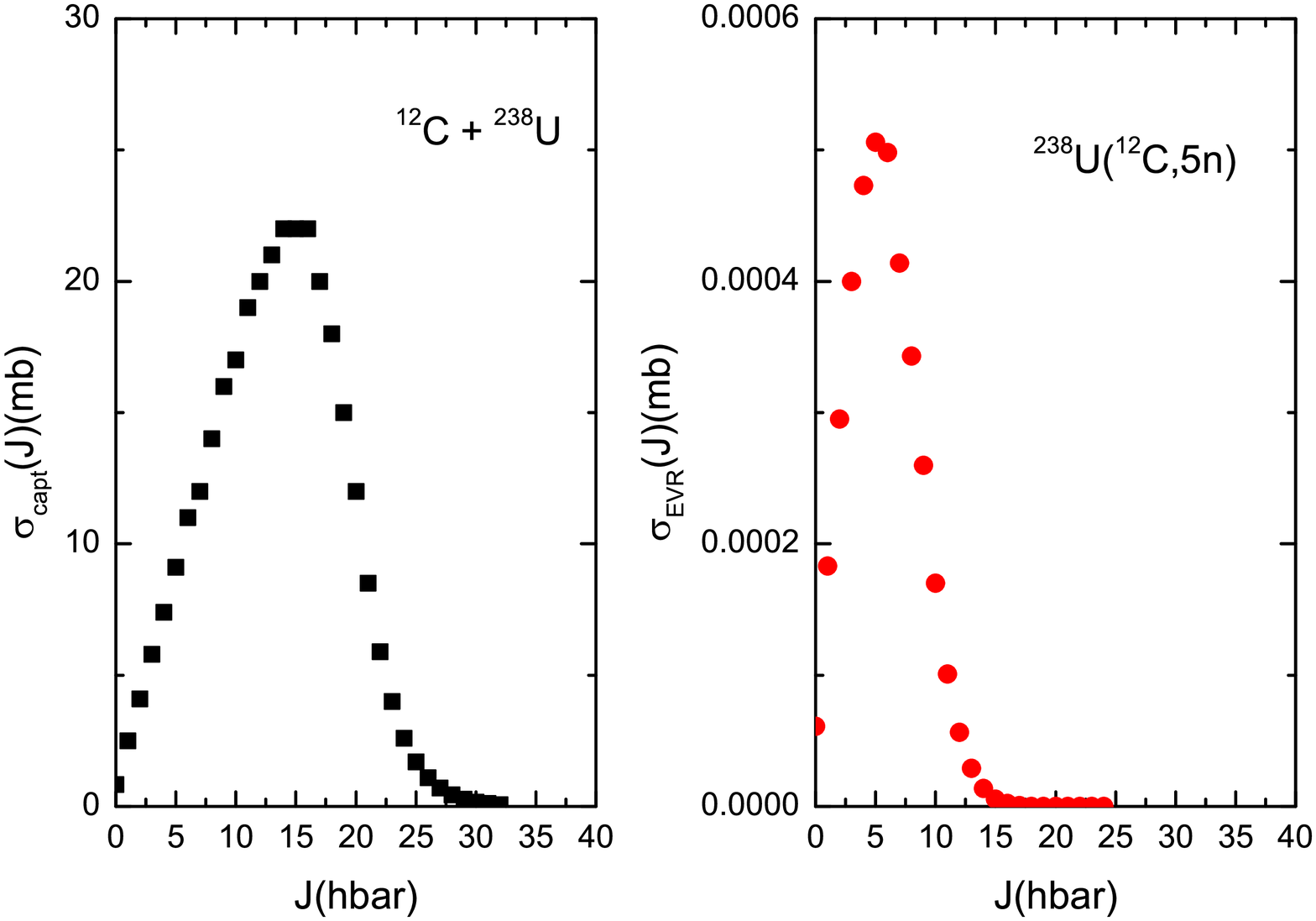}
\includegraphics[width=8cm]{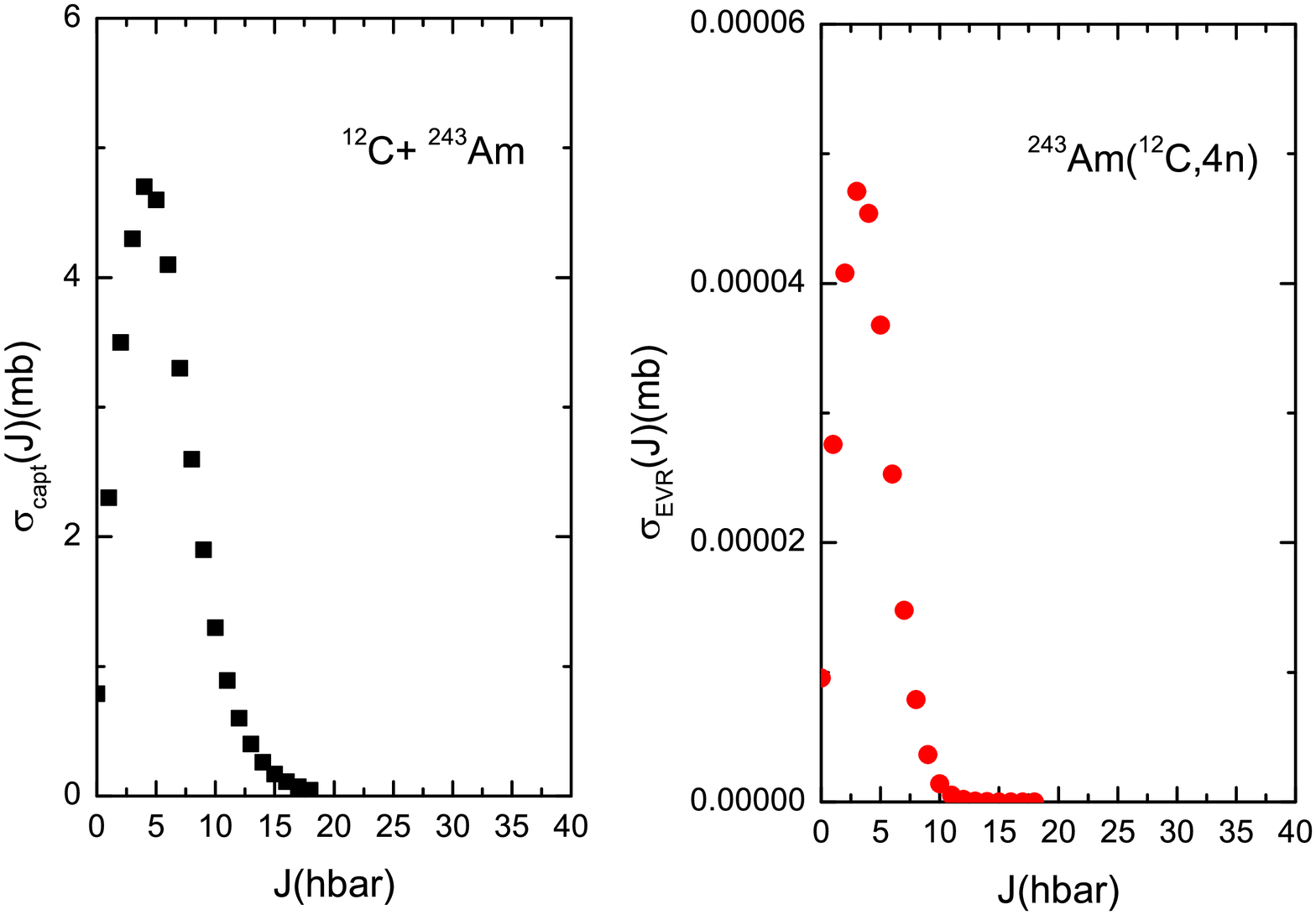}
\includegraphics[width=8cm]{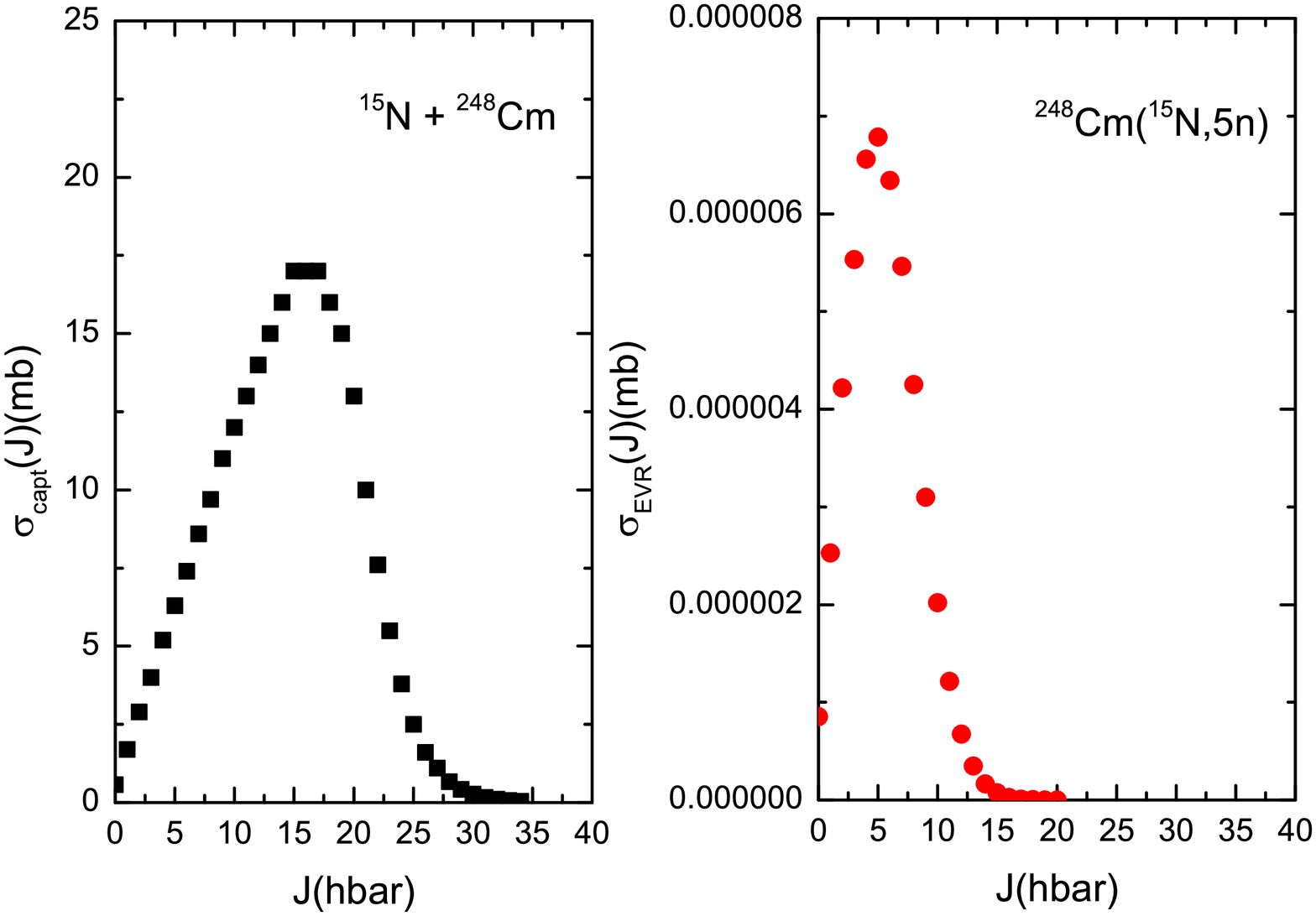}
\caption{Spin dependence of calculated capture cross sections and the EVR cross sections for the reactions of $^{12}$C + $^{232}$Th, $^{243}$Am, $^{249}$Cf and the reaction of $^{15}$N with $^{248}$Cm.  \cite{x,y,z,w} where the lab frame beam energies were 70,73,70, and 86 MeV respectively.}
\label{fig-3}       
\end{figure}

The most probable spins for the surviving evaporation residues are $\sim$ 5 $\hbar$, while the capture cross sections can involve spins out to 30 $\hbar$ with average spins of 10-20 $\hbar$.  Also note the dramatic reduction in cross section due to the fission of the completely fused system.  Capture cross sections for these hot fusion reactions are typically 100-300 mb while the evaporation residue cross sections are $\sim$ nb to $\mu$b.

For ``convenience" it is sometimes assumed that the relevant capture cross sections and P$_{CN}$ factors can be evaluated for J=0, i.e., drop the spin dependences.  This extreme view is not supported by the data shown in Fig.3.

\subsection{700 < Z$_{1}$Z$_{2}$ < 1000}

There are 119 cases in this category.  Many of these reactions are sub-barrier and the difference between the spin dependence of the  capture cross section and the spin-mediated evaporation residue cross section is smaller.  For all the cases studied, P$_{CN}$ =1.  A sampling of these cases is shown in Figure 4.  Note not only the decrease in the mean spin of the evaporation residues but the loss of orders of magnitude in the cross sections due to fission de-excitation.

\begin{figure}[h]
\centering
\includegraphics[width=8cm]{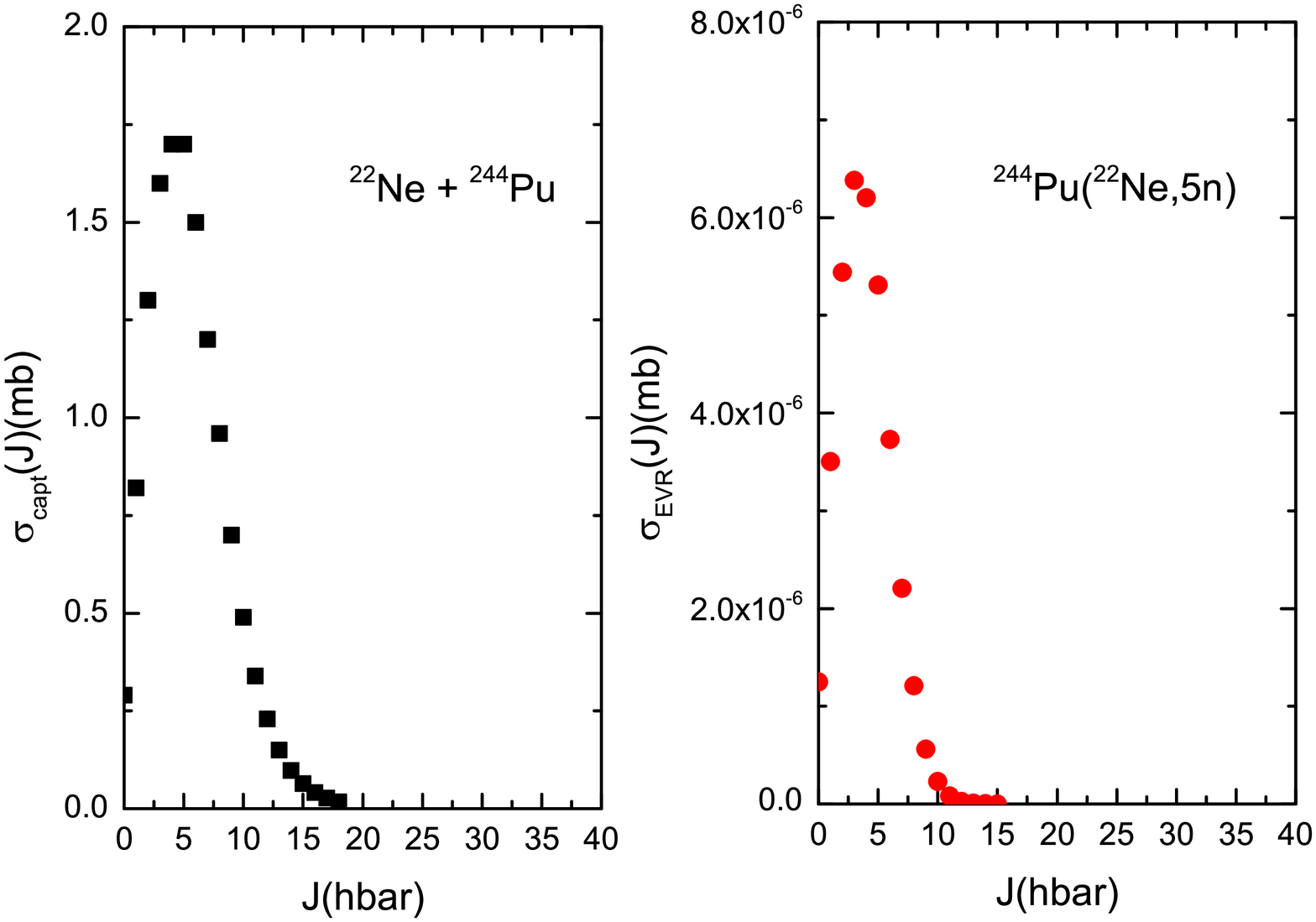}
\includegraphics[width=8cm]{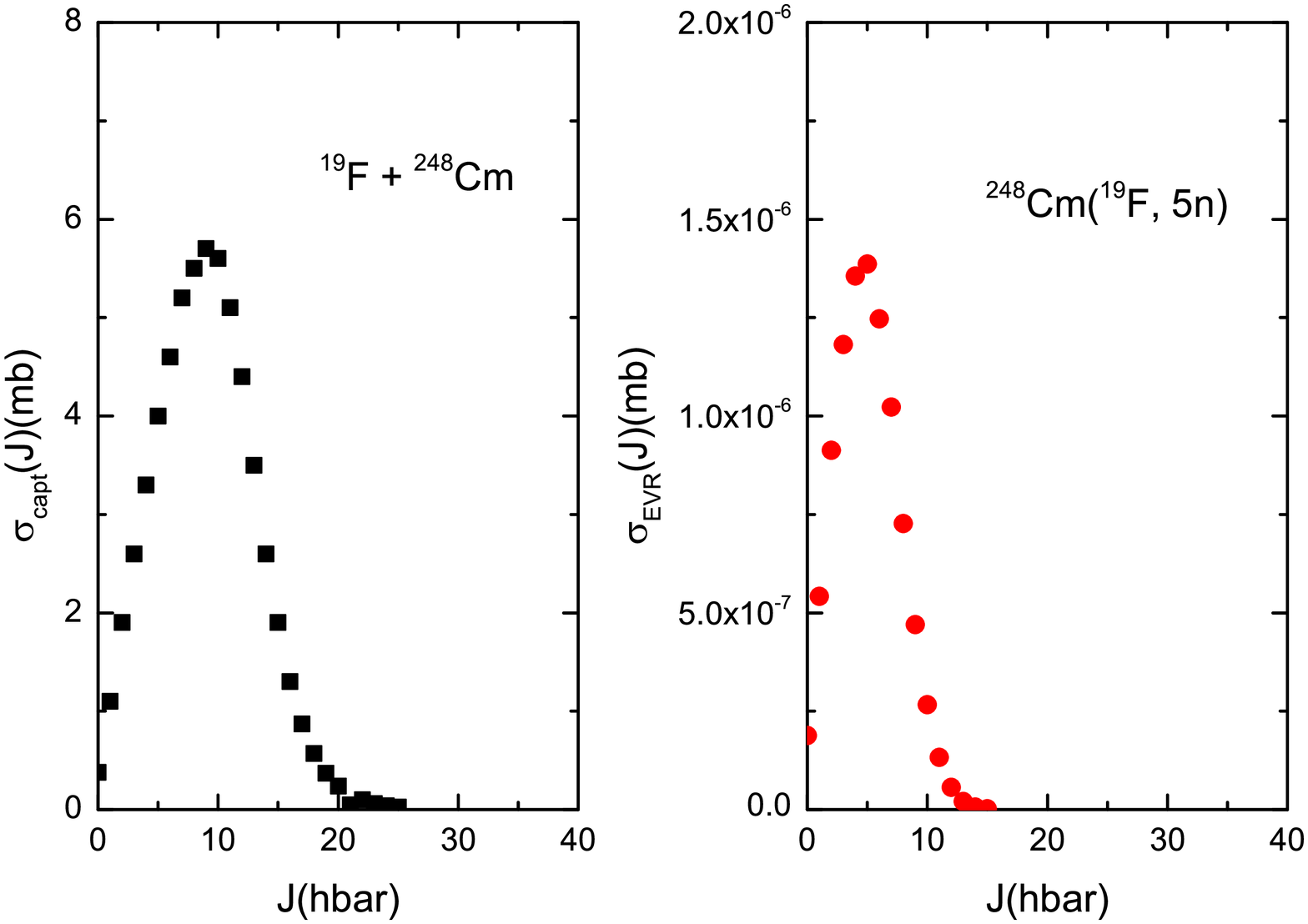}
\includegraphics[width=8cm]{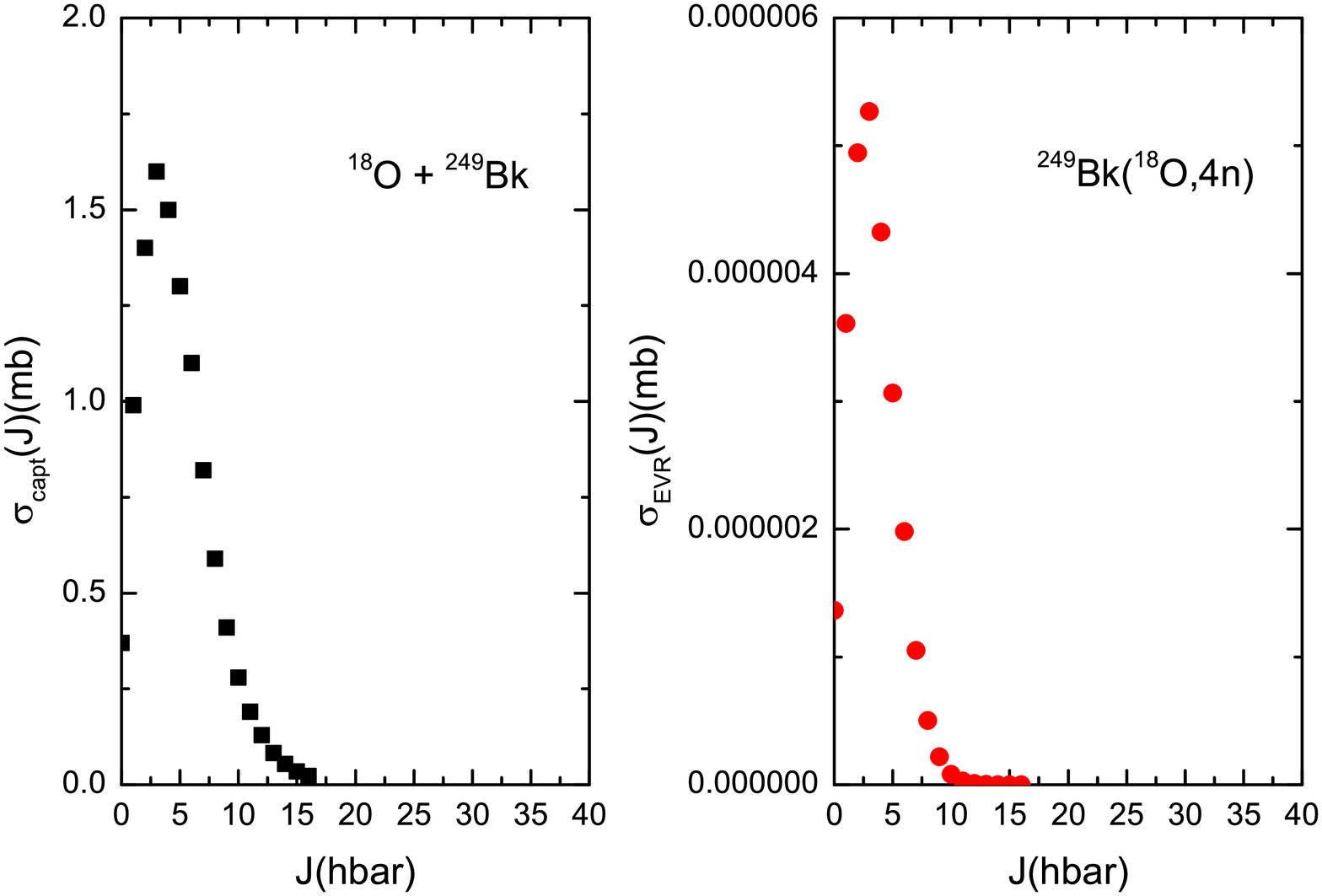}
\caption{Spin dependence of calculated capture cross sections and the EVR cross sections for the reaction of  $^{22}$Ne + $^{244}$Pu, $^{19}$F + $^{248}$Cm and $^{18}$O + $^{249}$Bk.  \cite{t,u,v} where the laboratory frame beam energies were 114,106, and 93  MeV, respectively.}
\label{fig-4}       
\end{figure}

\subsection{Z$_{1}$Z$_{2}$ > 1000}

We can break this category into two groups, 1000 $\leq$ Z$_{1}$Z$_{2}$ $\leq$ 1600 (136 cases) and Z$_{1}$Z$_{2}$ $\geq$ 1600 (167 cases).  Generally P$_{CN}$  $\sim$ 1 for the first group and P$_{CN}$ < 1 for Z$_{1}$Z$_{2}$ $\geq$ 1600.  In Figure 5, we show some typical cases of ``cold fusion" reactions with Z$_{1}$Z$_{2}$ $\geq$ 1600.  We plot the capture cross section and the ``effective" capture cross section for each case where the ``effective" capture cross section reflects the fraction of the capture cross section that contributes to evaporation residue formation.

\begin{figure}[h]
\centering
\includegraphics[width=8cm]{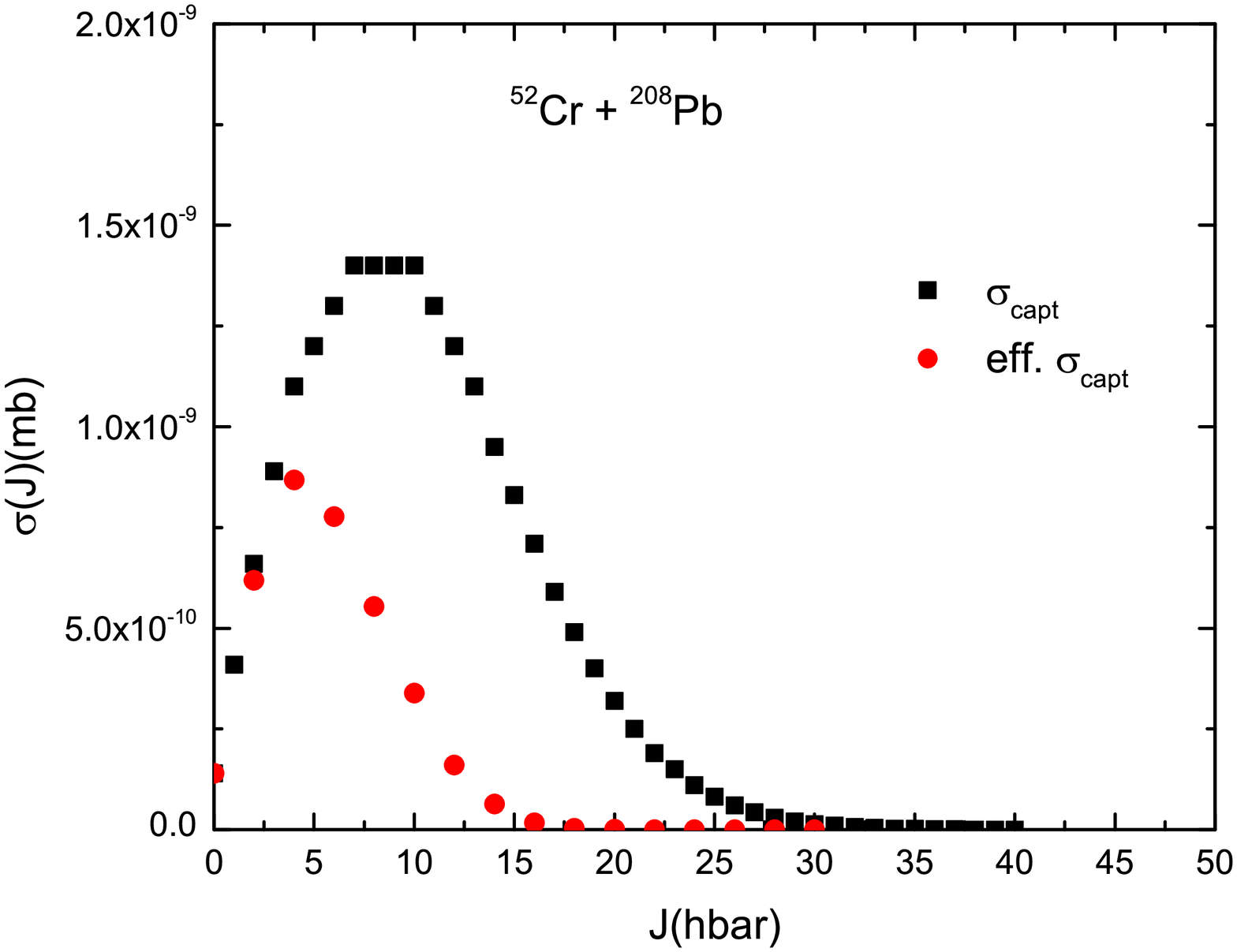}
\includegraphics[width=8cm]{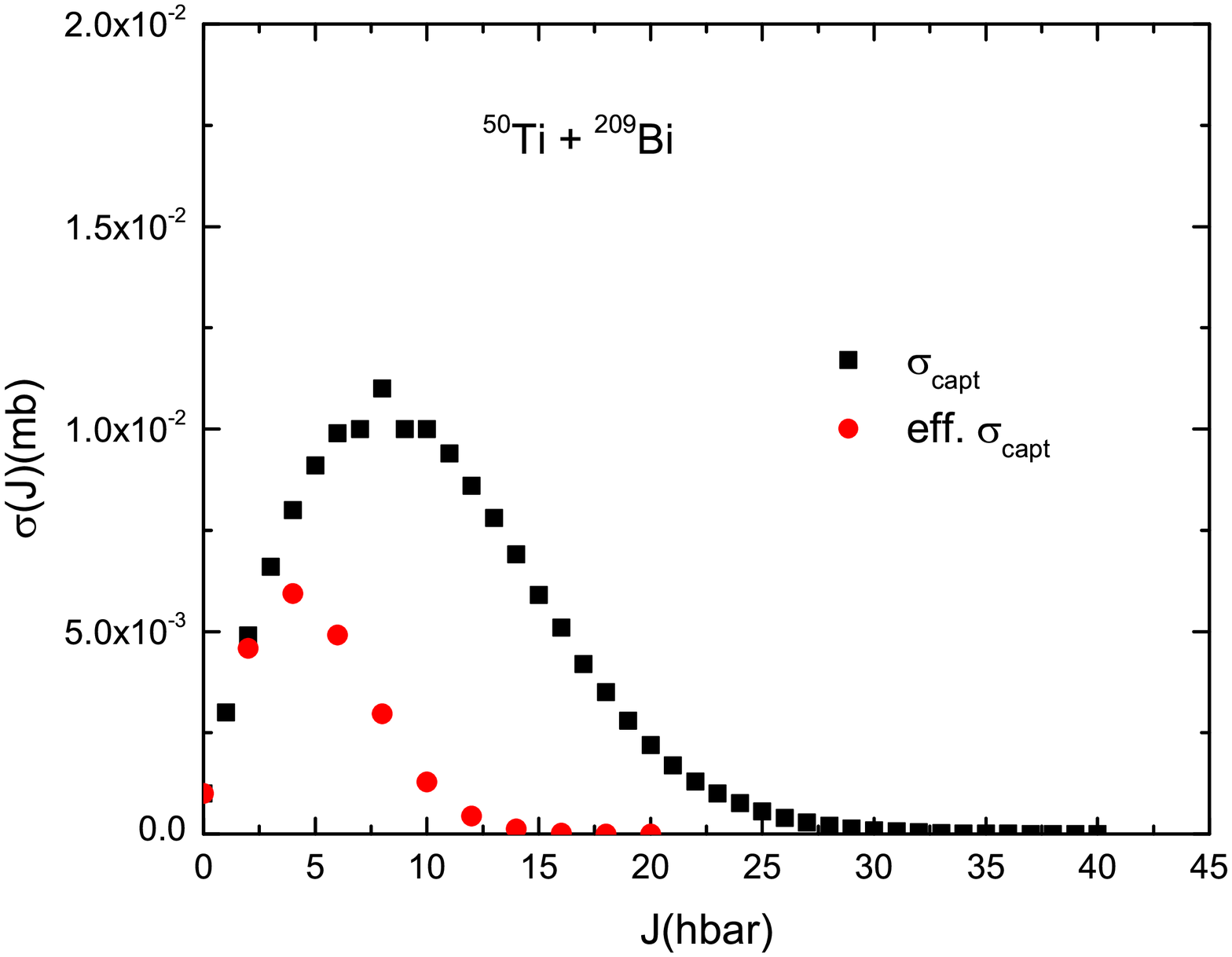}
\includegraphics[width=8cm]{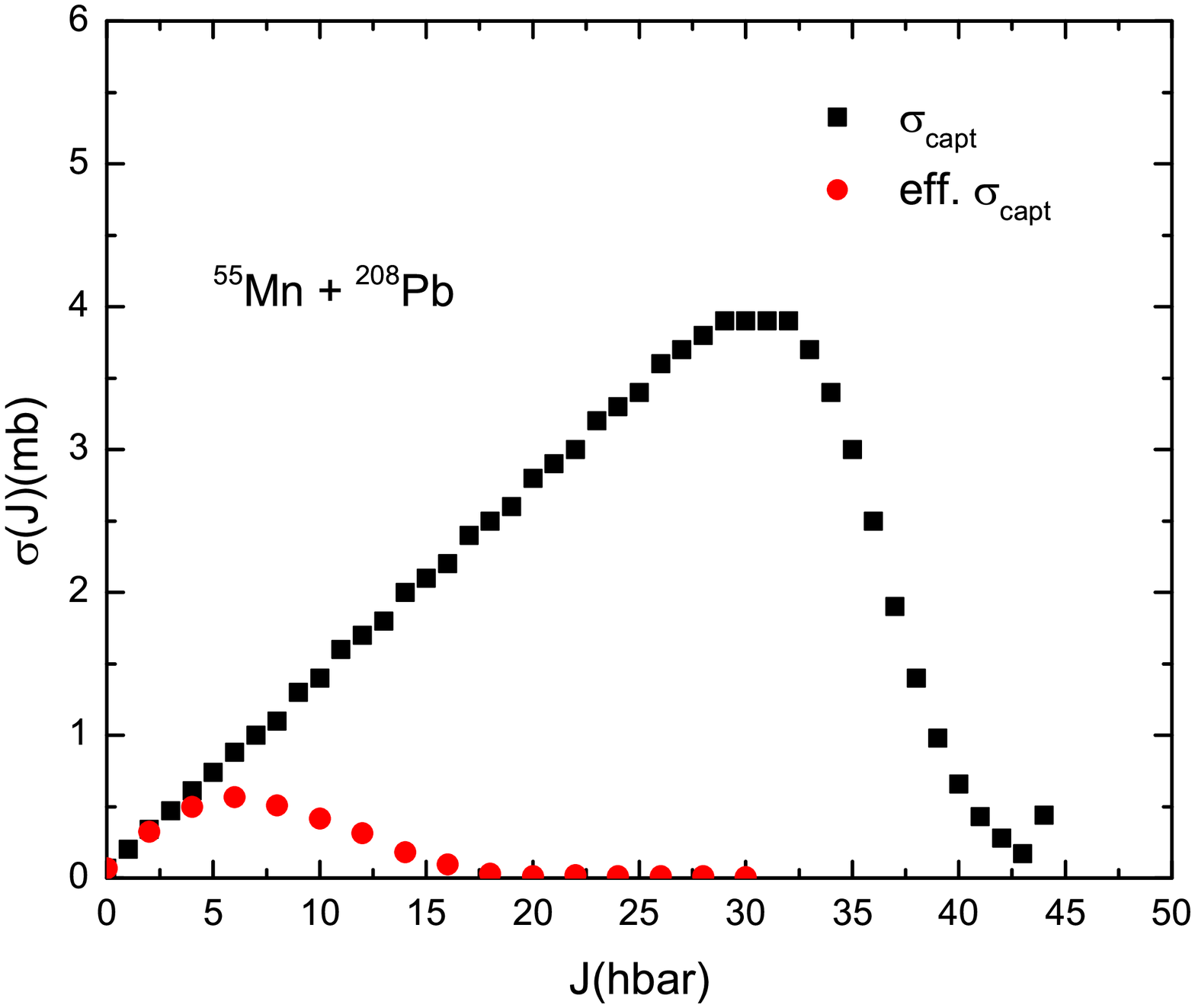}
\includegraphics[width=8cm]{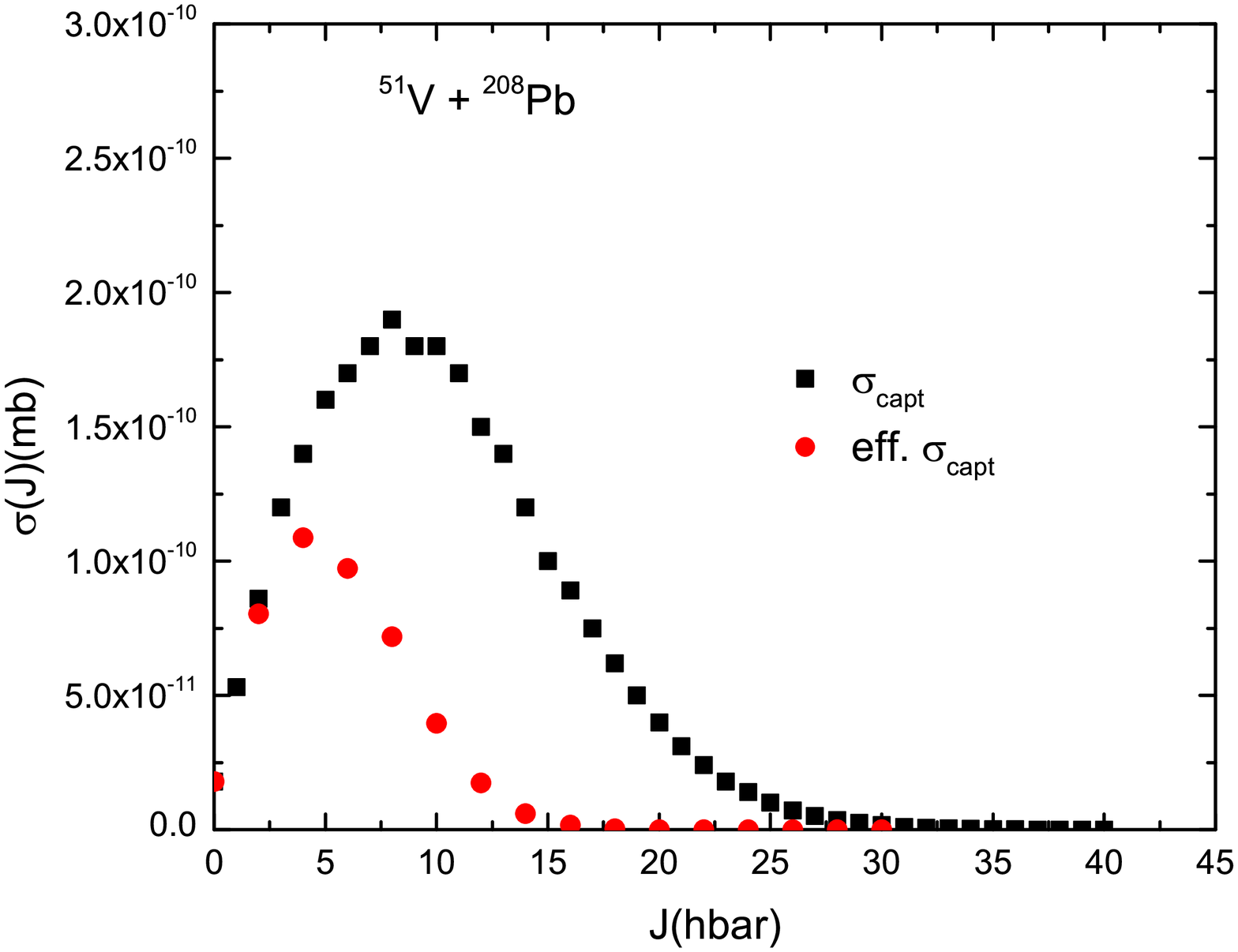}
\caption{Spin dependence of calculated capture cross sections and the ``effective" capture cross sections for the ``cold fusion"  reactions  \cite{m,n,o,p} where the lab frame beam energies were 253,241,283,and 240 MeV, respectively.}
\label{fig-5}       
\end{figure}

In Figure 6, we show a similar sampling of the data for hot fusion reactions.

\begin{figure}[h]
\centering
\includegraphics[width=8cm]{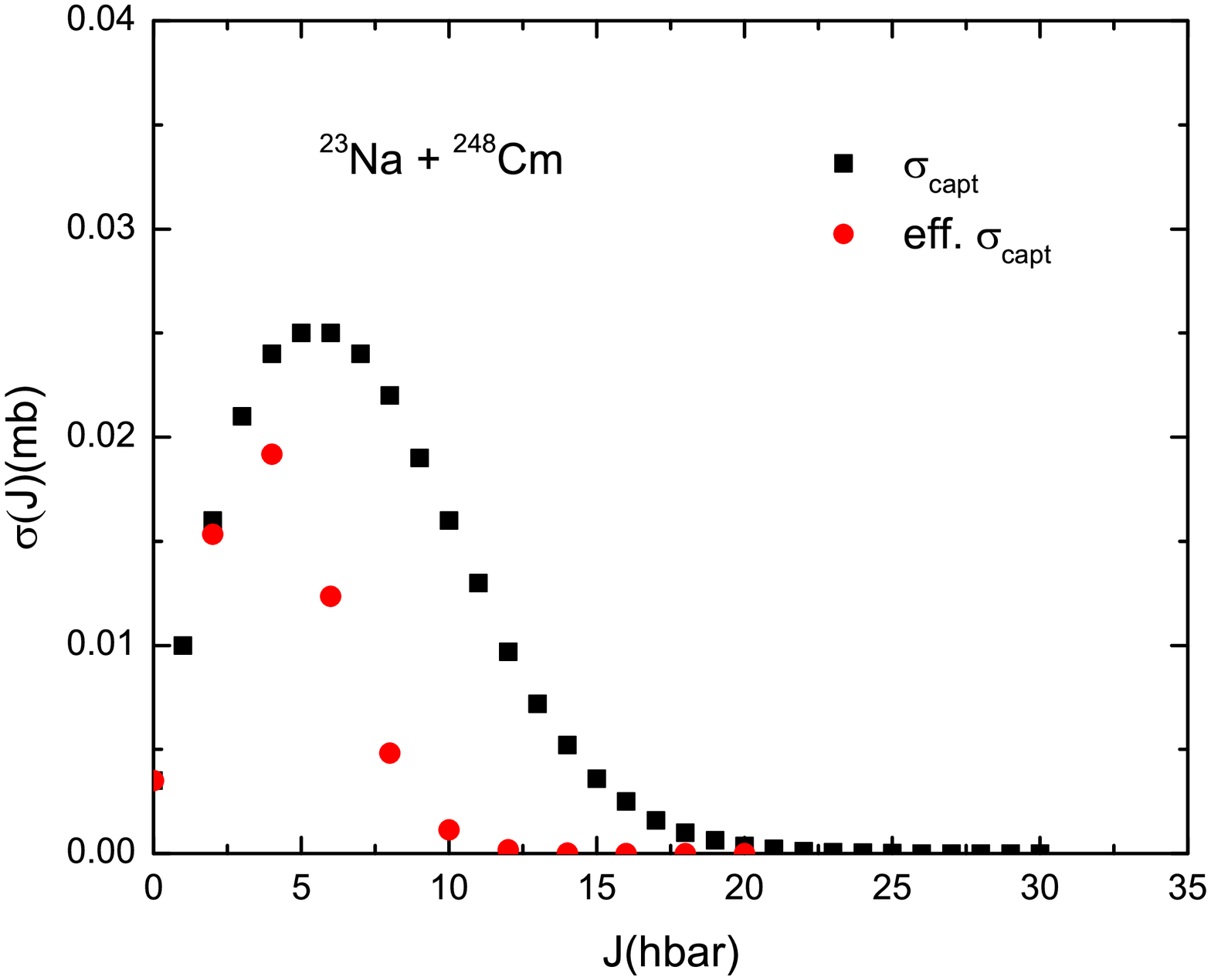}
\includegraphics[width=8cm]{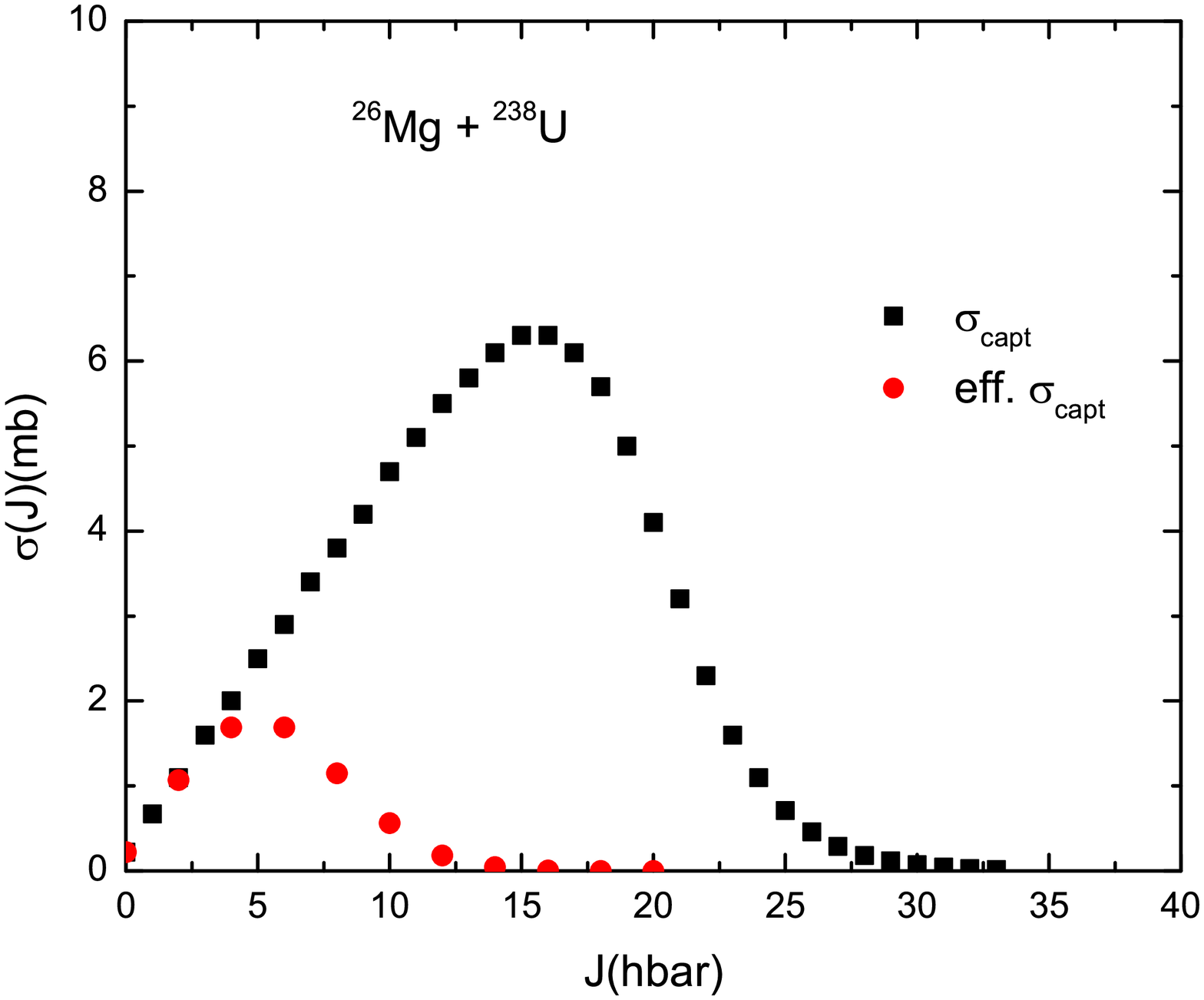}
\includegraphics[width=8cm]{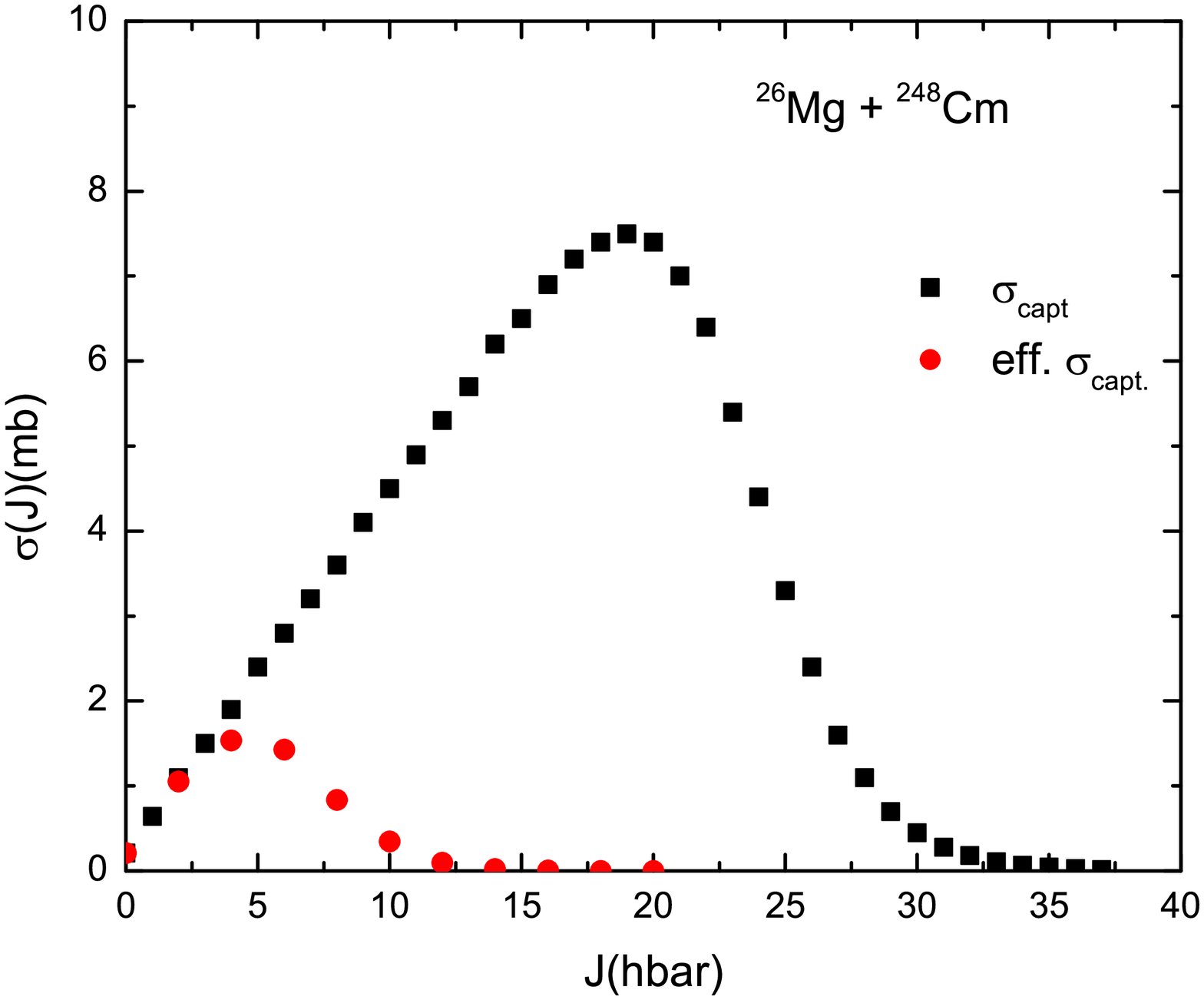}
\includegraphics[width=8cm]{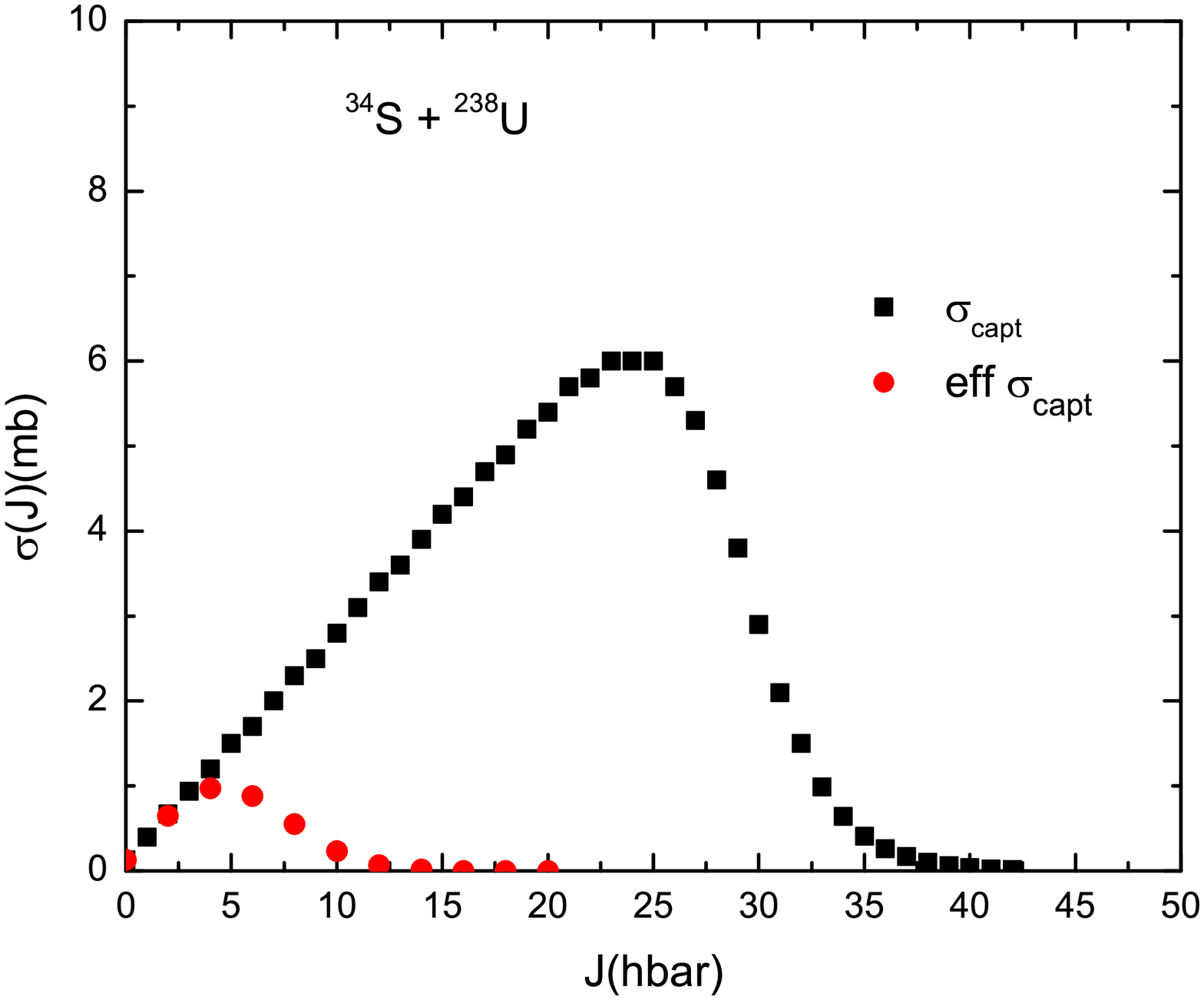}
\caption{Spin dependence of calculated capture cross sections and the ``effective" capture cross sections for the ``hot fusion"  reactions  \cite{f,g,h,i} where the lab frame beam energies were 130,139,144 and 188 MeV respectively.  }
\label{fig-6}       
\end{figure}

\subsection {Effective capture cross sections}

In Figure 7, we show the dependence of the ``effective" capture cross sections calculated in this work as a function of the product of the atomic numbers of the colliding nuclei, Z$_{1}$ and Z$_{2}$.  (Other scaling parameters could have been chosen such as (Z$_{1}$Z$_{2}$/(A$_{1}$$^{1/3}+$A$_{2}$$^{1/3}$) 
that might have been more effective but we have chosen the simplest parameterization).  One notes that the effective capture cross sections quickly decrease to less than 10 mb for Z$_{1}$Z$_{2}$ values > 1000 and for the most ``interesting" reactions (Z$_{1}$Z$_{2}$ > 1500) are only a few mb.  In terms of impact parameters for heavy element synthesis reactions, this means that the interesting collisions are the near central collisions.  Calculations of fusion and quasi-fission cross sections should focus on these collisions.

\section{Conclusions}

While measurements of capture, fusion and quasi-fission cross sections have scientific value, they must be ``filtered' for relevance in considerations of heavy element synthesis.  The ``effective" capture cross sections are a fraction of the ``real" capture capture cross sections due to the restrictions placed on these quantities due to the spin dependence of the survival probabilities.  The ``effective" capture cross sections decrease rather than increase with increasing Z$_{1}$Z$_{2}$ due to these survival probability effects.  Only a few partial waves contribute to the effective capture cross sections for larger values of Z$_{1}$Z$_{2}$. Calculations of fusion cross sections for the synthesis of the heaviest nuclei should only consider small impact parameter collisions.

\begin{figure}[h]
\centering
\includegraphics[width=9cm]{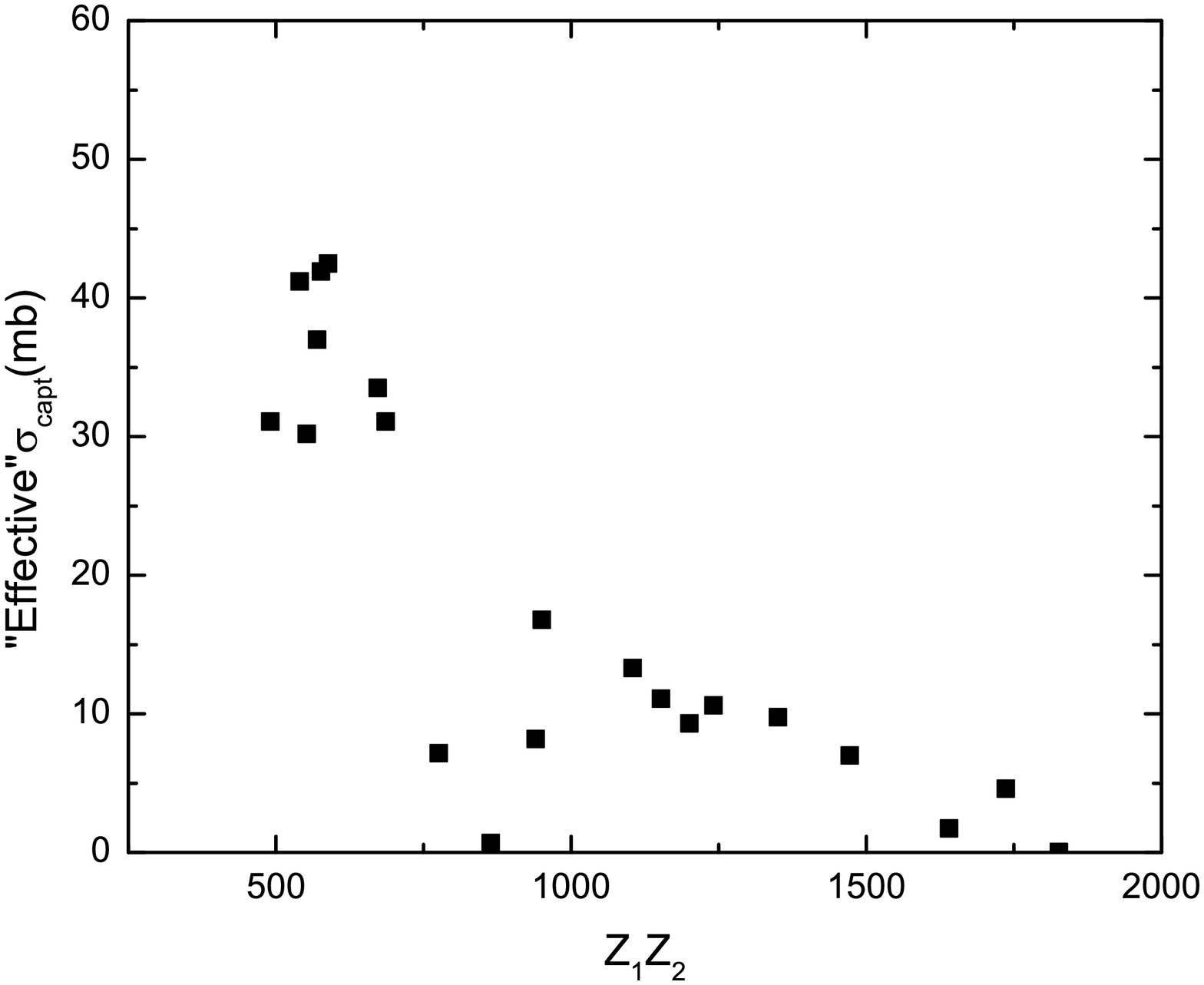}
\caption{The dependence of the effective capture cross sections calculated in this work on the scaling parameter Z$_{1}$Z$_{2}$}
\label{fig-7}       
\end{figure}

\section {Acknowledgements}

This work was supported in part by the U.S. Dept. of Energy, Office of Science, Office of Nuclear Physics under award number DE-SC0014380. 
%

\begin{thebibliography}{}
%
%
\bibitem{Carolla}
C. Laue, et al., , Phys. Rev. C  \textbf{59}, 3086 (1999)

\bibitem{Zaggy}

Nuclear Reactions Video Project (nrv.jinr.ru) -``Statistical Model of Decay of Excited Nuclei"

\bibitem{Ig83}

A. V. Ignatyuk, IAEA report INDC(CCP)-233/L(1985)

\bibitem{Mol16}

P. Moller, A.J. Sierk, T. Ichikawa, and H. Sagawa, At. Data and Nucl. Data Table \textbf{109-110}, 1 (2016).

\bibitem{Zaggy01}

V.I. Zagrebaev, Y. Aritomo, M.G. Itkis, Y.T. Oganessian, and M. Ohta, Phys. Rev. C  \textbf{65}, 014607 (2002)

\bibitem{Kr40}

H.A. Kramers, Physica (Amsterdam) \textbf{7}, 284 (1940).

\bibitem{lb}

H. Lu, D. Boilley, EPJ Web of Conferences {\bf 62},03002 (2013).

\bibitem{wdl}

W. Loveland, Eur. J. Phys. A \textbf{51}, 120 (2015).

\bibitem{Chris}

Ch. E. Duellmann, private communication

\bibitem{oleg}

R. Bass, Phys. Rev. Lett. \textbf{39}, 265 (1977); R. Bass, Nuclear reactions with heavy ions (Springer-Verlag, Berlin, 1980).

\bibitem{greg}

G. Henning, et al., Phys. Rev. Lett. \textbf{113}, 262505 (2014).

\bibitem{x}

Z. Qin et al. , Radiochemica Acta, \textbf{96}, 455 (2008)

\bibitem{y}

 T. Sikkeland, J. Maly, and D.F. Lebeck,  Phys. Rev. \textbf{169}, 1000 (1968)
\bibitem{z}

P. Eskola, Phys. Rev. C \textbf{7}, 280 (1973)

\bibitem{w}

K. Eskola, P. Eskola, M. Nurmia and A. Ghiorso,  Phys. Rev. C \textbf{4},  632 (1971)



\bibitem{t}

Yu. A. Lazarev et al. , Phys. Rev. C \textbf{62}, 064307 (2000)


\bibitem{u}

Y. Nagame, et al., J. Nucl. Radio Sciences \textbf{3}, 85 (2002).


\bibitem{v}

J. V. Kratz et al., Phys. Rev, C \textbf{45}, 1064 (1992).

\bibitem{m}

C. M. Folden III et al., Phys. Rev. C \textbf{79}, 027602  (2009)

\bibitem{n}

F.P. Hessberger et al., Eur. Phys. J. A \textbf{12}, 57 (2001)

\bibitem{o}


S. L. Nelson et al., Phys. Rev.  C \textbf{78}, 024606 (2008)


\bibitem{p}

J.M. Gates et al., Phys. Rev. C \textbf{78}, 034604 (2008)


\bibitem{f}

K. Morita. et al., J. Phys. Soc. Japan \textbf{78}, 064201 (2009).


\bibitem{g}

J.M. Gates et al., Phys. Rev. C \textbf{77}, 034603  (2008)

\bibitem{h}

J. Dvorak et al., Phys. Rev. Lett. \textbf{100}, 132503  (2008)



\bibitem{i}



K. Nishio et al., Phys. Rev. C \textbf{82}, 024611 (2010)






\end{thebibliography}
%
%

\end{document}